\documentclass[manuscript]{aastex}
\usepackage{natbib}
\usepackage{gensymb}
\usepackage{color}
\newcommand{\Msun}{M$_{\odot}$}

\slugcomment{ApJL in press (DOI: 10.3847/2041-8213/aa8edf )}
\shorttitle{The discovery of the electromagnetic counterpart of GW170817}
\shortauthors{Valenti, Sand, Yang et al.}

\begin{document}
 \title{The discovery of the electromagnetic counterpart of GW170817: kilonova AT~2017gfo/DLT17ck}

\author{Stefano Valenti,$\!$\altaffilmark{1} David, J. Sand,$\!$\altaffilmark{2} Sheng Yang,$\!$\altaffilmark{1,3}
\and
 Enrico Cappellaro,$\!$\altaffilmark{3} Leonardo Tartaglia,$\!$\altaffilmark{1,2} 
 \and
 Alessandra Corsi,$\!$\altaffilmark{4}  Saurabh W. Jha,$\!$\altaffilmark{5}
Daniel E. Reichart,$\!$\altaffilmark{6} Joshua Haislip,$\!$\altaffilmark{6} Vladimir Kouprianov,$\!$\altaffilmark{6}}

\begin{abstract}
During the second observing run of the Laser Interferometer gravitational-wave Observatory (LIGO) and Virgo Interferometer,  a gravitational-wave signal consistent with a binary neutron star coalescence was detected on 2017 August 17th (GW170817), quickly followed by a coincident short gamma-ray burst trigger by the \textit{Fermi} satellite. The Distance Less Than 40 (DLT40) Mpc supernova search performed pointed follow-up observations of a sample of galaxies regularly monitored by the survey which fell within the combined LIGO+Virgo localization region, and the larger  \textit{Fermi} gamma ray burst error box.  Here we report the discovery of a new optical transient (DLT17ck, also known as SSS17a; it has also been registered as AT~2017gfo) spatially and temporally coincident with GW170817. The photometric and spectroscopic evolution of DLT17ck are unique, with an absolute peak magnitude of $M_{r}$ = -15.8 $\pm$ 0.1 and an $r-$band decline rate of $1.1\,\rm{mag}/\rm{d}$.  This fast evolution is generically consistent with kilonova models, which have been predicted as the optical counterpart to binary neutron star coalescences. Analysis of archival DLT40 data do not show any sign of transient activity at the location of DLT17ck down to $r$\/$\sim$19 mag in the time period between 8 months and 21 days prior to GW170817. This discovery represents the beginning of a new era for multi-messenger astronomy opening a new path to study and understand binary neutron star coalescences, short gamma-ray bursts and their optical counterparts.
\end{abstract}

 
\altaffiltext{1}{Department of Physics, University of California, 1 Shields Avenue, Davis, CA 95616-5270, USA}
\altaffiltext{2}{Department of Astronomy/Steward Observatory, 933 North Cherry Avenue, Room N204, Tucson, AZ 85721-0065, USA}
\altaffiltext{3}{INAF Osservatorio Astronomico di Padova, Vicolo dell’Osservatorio 5, I-35122 Padova, Italy}
\altaffiltext{4}{Physics $\&$ Astronomy Department, Texas Tech University, Lubbock, TX 79409, USA 0000-0003-3433-1492}
\altaffiltext{5}{Department  of  Physics  and  Astronomy,  Rutgers,  TheState University of New Jersey, Piscataway, NJ 08854, USA}
\altaffiltext{6}{Department of Physics and Astronomy, University of North Carolina at Chapel Hill, Chapel Hill, NC 27599, USA}

\section{Introduction}

The era of multi-messenger astronomy has truly begun.  During the first Advanced LIGO \citep[aLIGO, ][]{Aasi2015} run (O1), two definitive gravitational wave (GW) events were observed, corresponding to relatively massive black hole -- black hole (BH-BH) mergers of 36+25 $M_{\odot}$ \citep[GW150914; ][]{GW150914} and 14+8 $M_{\odot}$ \citep[GW151226; ][]{GW151226}, respectively. These amazing discoveries were followed by a third event during the second aLIGO run (O2), which was another massive BH-BH merger \citep[31+19 $M_{\odot}$]{GW170104}.  Each GW event was accompanied by a massive effort from the astronomical community to identify an electromagnetic (EM) counterpart \citep[see e.g.][]{GWfollowup}, even though the likelihood of finding EM counterparts to BH-BH mergers is low.  On the other hand, GW events including at least one neutron star (NS; as either a NS-NS or NS-BH coalescence) are expected to produce a variety of EM signatures. Chief among them in the optical+near infrared regime is the so-called kilonova, resulting from the decay of r-process elements produced and ejected during the merger process  \citep[for a review, see][]{Metzger17}.

In order to search for kilonovae, two general approaches have been proposed: 1) wide-field searches of the aLIGO localization region \citep[see e.g. ][]{2016ApJ...827L..40S} and 2) narrow-field targeted searches of galaxies both at the predicted GW event distance and within the sky localization region \citep[e.g.][]{2016ApJ...820..136G}. The $D$\/$<$40 (DLT40) Mpc one-day cadence supernova search uses the second approach, targeting galaxies within the GW localization region which are part of the main supernova search. 

On 2017 August 17 (UT) a GW event was discovered  by aLIGO and Virgo \citep{Acernese2015} observatories (LIGO-Virgo collaboration, hereafter LVC) which was consistent with a neutron star binary coalescence with a low false alarm rate, at a distance of $D\sim40\,\rm{Mpc}$ \citep[][see Section 3 for further details]{2017GCN..21506....1A}. A potential short gamma-ray burst (GRB) counterpart was discovered by \textit{Fermi} (GBM; trigger 524666471). Soon after, multiple groups reported the detection of an optical counterpart (AT~2017gfo/DLT17ck/SSS17a; we will refer to the counterpart as DLT17ck throughout this work) which was subsequently identified as a true kilonova based on its fast spectroscopic \citep{2017GCN..21582....1A} and photometric \citep{2017GCN..21579....1A} evolution.  The apparent host galaxy of DLT17ck is the normal early-type galaxy NGC~4993 \citep[see][]{2017GCN..21645....1A}.

In this paper we present the observations of the DLT40 team associated with the kilonova DLT17ck, based on our ongoing one-day cadence search. The DLT40 team was one of the initial groups reporting the discovery of the kilonova (Section~\ref{sec:discovery}), and based on our light curve and an early spectrum, we show that DLT17ck resembles the expected observables of a kilonova (Section~\ref{sec:dlt17ck}).  The DLT40 team also has a history of observations of the NGC~4993 field during the year before the GW event (Section~\ref{sec:prediscovery}).  We end the paper by summarizing our results, and discussing prospects for electromagnetic counterpart searches with small telescopes (Section~\ref{sec:summary}).

\section{DLT40 GW counterpart search}\label{sec:search}

The Distance Less than 40 Mpc survey (DLT40;  Tartaglia et al., in preparation) is a one day cadence supernova search using a PROMPT 0.4m telescope located at Cerro Tololo Inter-American Observatory  \citep[CTIO; ][]{PROMPT}.  The survey goal is the early detection and characterization of nearby SNe.   DLT40 has been operational since 2016, and observes $\sim$300--600 targeted galaxies on a nightly basis. A typical single-epoch integration of 45 s reaches a limiting magnitude of $r\approx19\,\rm{mag}$ with filterless observations.  The field of view of the PROMPT camera is 10$\times$10 arcmin$^2$, sufficient to map all but the nearest galaxies in the search.

The DLT40 galaxy sample is drawn from the Gravitational Wave Galaxy Catalogue \citep[GWGC; ][]{GWGC}, with further cuts made on recessional velocity ($V$\/$<$ 3000 km/s, corresponding to $D$\/$\lesssim$40 Mpc), declination (Dec$<$+20 deg), absolute magnitude ($M_{B}$\/$<$\/$-$18 mag), and Milky Way extinction ($A_V$\/$<$0.5 mag). For these galaxies, we strive for a one-day cadence between observations to constrain the explosion epoch of any potential SN.

The DLT40 pipeline is totally automated, with pre-reduced images delivered from the telescope, ingested and processed in a few minutes.
New transient candidates are detected on difference images and are available for visual inspection within $\sim$2-3 minutes of ingestion.
 At the time of writing, DLT40 has discovered and confirmed 12 young SNe in the nearby Universe, with initial results reported in \citet{Hosseinzadeh17}.

The DLT40 GW follow-up strategy was planned as a straightforward addition to the core DLT40 supernova search.  When a LVC gravitational wave event is announced, the {\sc BAYESTAR} sky map \citep{Singer16} is cross-matched with the DLT40 galaxy sample. All DLT40 galaxies that are within the LVC localization area are selected for high priority imaging in the nightly DLT40 schedule. Depending on the size of the LVC localization area and the number of galaxies selected,  we apply a spatial cut (between the 80-99$\%$ confidence localization contours of the LVC map) and/or a cut  in luminosity to select the galaxies with most of the stellar mass.  This strategy broadly follows that laid out by other LVC EM follow-up groups with narrow-field telescopes \citep[e.g.][]{2016ApJ...820..136G}. Further details of our search strategy, and our other results from O2, will be presented in a separate work (Yang et al. in preparation).  

\section{Discovery of DLT17ck}
\label{sec:discovery}
\begin{figure*}
\begin{center}
\includegraphics[width=16cm]{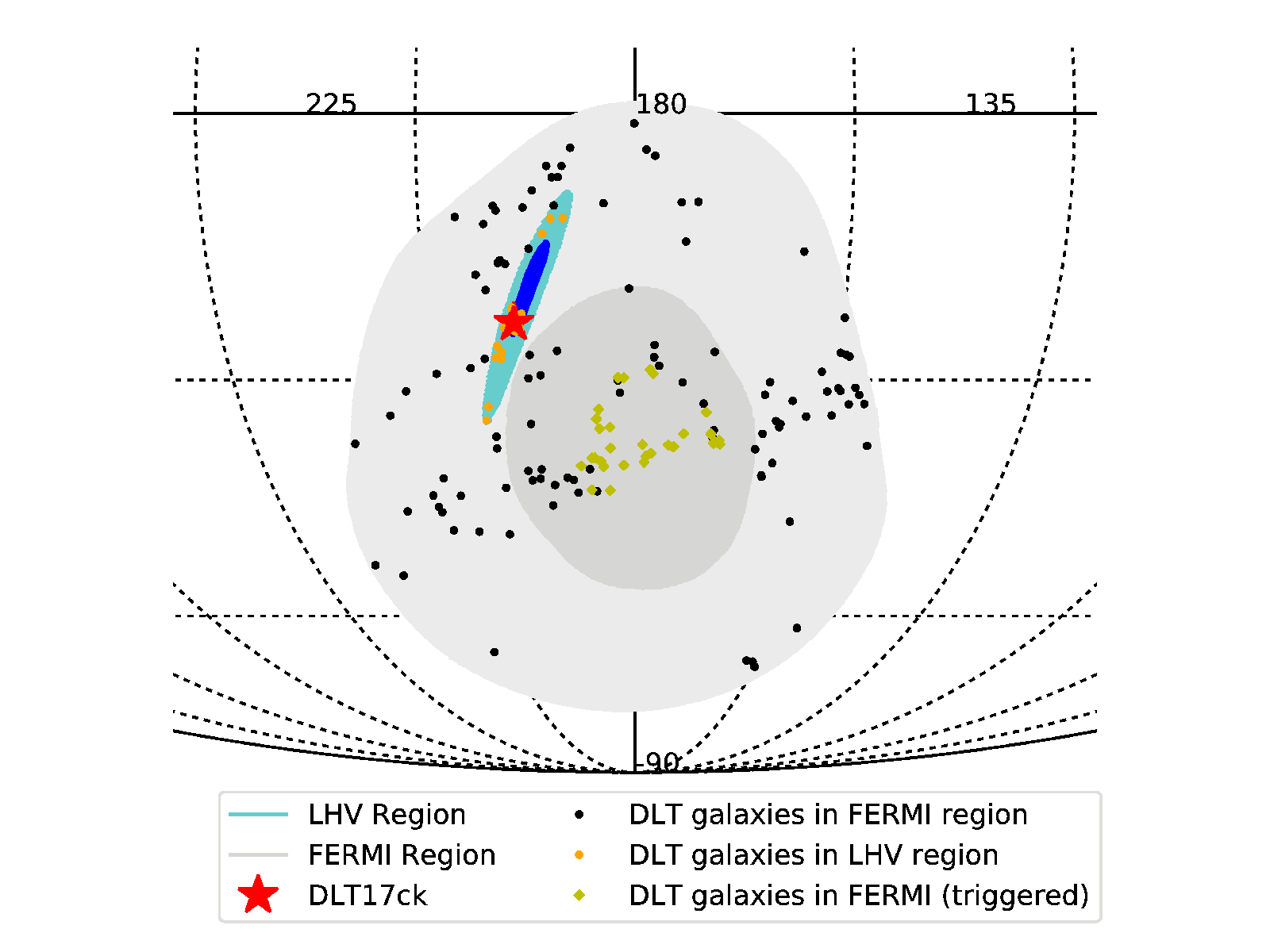}
\caption{The sky map region of the GW170817 LVC event using all three gravitational-wave observatories (H1, L1, and V1)
over-imposed on the \textit{Fermi} localization of GBM trigger 524666471/170817529. The DLT40 galaxies observed the first Chilean night after the LVC trigger are marked in orange (galaxies within the LVC region) and in olive green (galaxies within the \textit{Fermi} localization).  The remaining black points are those DLT40 galaxies which were within the \textit{Fermi} localization but were not observed by our program.  The red star marks the location of DLT17ck and the host galaxy NGC~4993.}
\label{fig:ligoregion}
\end{center}
\end{figure*}

On 2017 August 17.528 UT, the LVC reported the detection of a gravitational-wave nearly coincident in time \citep[2 seconds before, ][]{2017GCN..21506....1A} with the \textit{Fermi} GBM trigger 524666471/170817529 located at RA=176.8$^{\circ}$ and DEC=-39.8$^{\circ}$ with an error of 11.6$^{\circ}$ (at 1$\sigma$). The LVC candidate  had an initial localization of RA=186.62$^{\circ}$,  DEC=$-$48.84$^{\circ}$  and a 1$\sigma$ error radius of 17.45$^{\circ}$  \citep{2017GCN..21505....1A}. 
The GW candidate was consistent with a neutron star binary coalescence with false alarm rate of $\sim$ 1/10,000 years \citep{2017GCN..21505....1A}. The gravitational wave was clearly detected in the LIGO detectors but was below threshold for the Virgo detector \citep{2017GCN..21513....1A}. Despite this, the Virgo data were still crucial to further constrain the localization of the event to only 
31 deg$^2$ (90\% credible region). The luminosity distance was constrained with LIGO data to be 40 $\pm$ 8 Mpc 
\citep{2017GCN..21513....1A}.  In Figure ~\ref{fig:ligoregion} we show a map of both the LIGO+Virgo and \textit{Fermi} GBM localizations, which overlapped on the sky.  As part of the DLT40 search, we prioritized observations of 20 galaxies within the 99$\%$ confidence area of the LVC error-box and with a cut in luminosity. Among the 23 galaxies within the LIGO/Virgo error box, we selected the 20 galaxies within 99$\%$ of the cumulative luminosity distribution. At the same time, we also selected the 31 most luminous galaxies in the \textit{Fermi} region of the coincident short GRB (see Figure \ref{fig:ligoregion}). The 51 DLT40 galaxies selected were then observed at high priority. 
In this work, we present the only transient we detected within either the LVC or \textit{Fermi} localizations: AT 2017gfo/DLT17ck (detected in in NGC~4993).

\begin{figure*}
\begin{center}
\includegraphics[width=16cm]{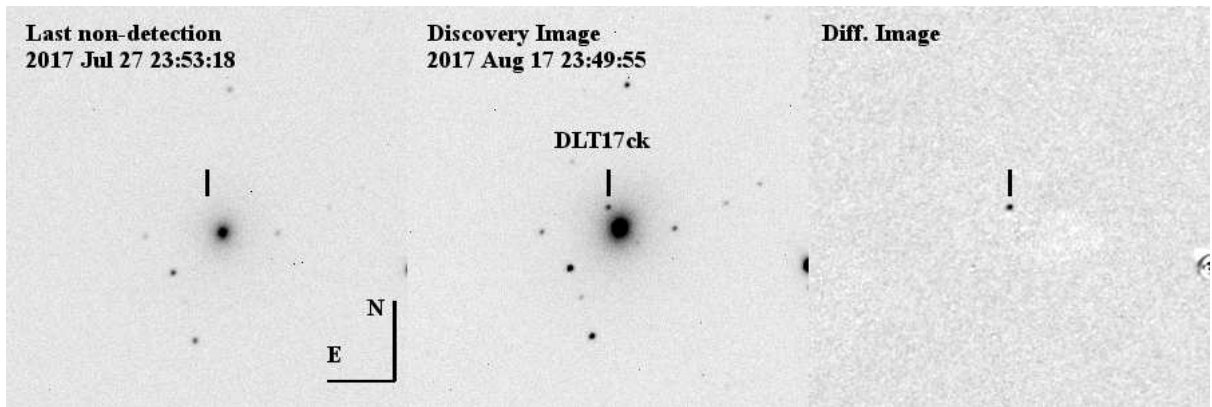}
\caption{Last non-detection (on the left), discovery image of DLT17ck observed on 2017-08-17 at 23:49:55 UT. The difference image  is shown on the right, where DLT17ck is clearly visible.}
\label{fig:FC}
\end{center}
\end{figure*}

 On 2017 August 17  23:49:55 UT (11.09 hours after the LVC event GW170817), we detected DLT17ck, at RA=13:09:48.09 and DEC=-23:22:53.4.6,  5.37W, 8.60S arcsec offset from the center of NGC~4993 \citep[][see Figure \ref{fig:FC}]{2017GCN..21531....1A}. At the same time, DLT17ck was detected by  \cite{2017GCN..21529....1A}, \cite{2017GCN..21530....1A}, \cite{2017GCN..21532....1A} and \cite{2017GCN..21565....1A} and intensively observed by 
 portions of the astronomical community that signed a memorandum of understanding with the LVC.  DLT17ck was not reported to the internal (collaboration-wide) GCN by our team until a second confirmation image was obtained on August 18 00:40:38 UT. The LVC GW region of GW170817 was also observed in other windows of the EM spectrum, from radio to X-ray wavelengths. It was recovered in the UV, optical, and near infrared. Deep X-ray follow-up observations conducted with the \textit{Chandra} observatory revealed X-ray emission from a point source at a position consistent with that of the optical transient DLT17ck \citep{2017GCN..21568....1A,2017GCN..21786....1A,2017GCN..21787....1A}. A radio source consistent with the position of DLT17ck \citep{2017GCN..21816....1A} was detected with the Karl G. Jansky VLA \citep{2017GCN..21814....1A,2017GCN..21815....1A}, at two different frequencies ($\approx 3$\,GHz  and $\approx 6$\,GHz). Marginal evidence for radio excess emission at the location of DLT17ck was also found in ATCA images of the field at similar radio frequencies  \citep[$\approx 5$\,GHz;][]{2017GCN..21568....1A}. Finally, neutrino observations were reported with one neutrino detected within the preliminary LVC localization \citep{2017GCN..21511....1A}, though this was later established to be unrelated to GW170817/DLT17ck \citep{2017GCN..21568....1A}.
 
\section{DLT17ck: a new type of transient} 
\label{sec:dlt17ck}
\begin{figure*}
\begin{center}
\includegraphics[width=16cm]{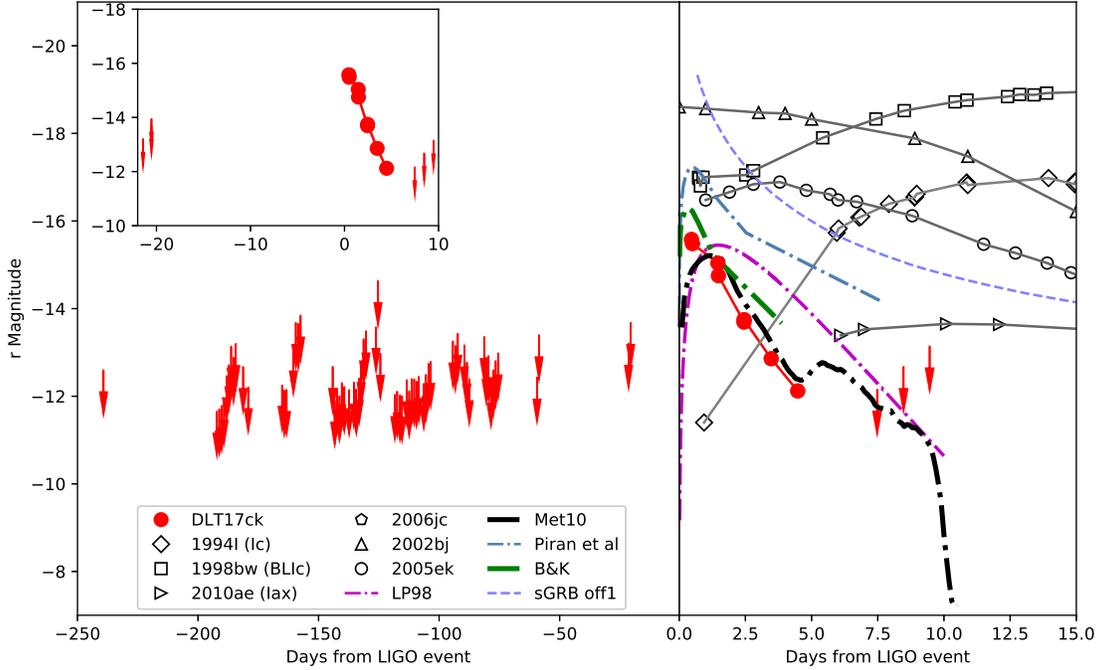}
\caption{{\bf Right panel:} DLT40 light curve of DLT17ck (in red) over plotted with normal or fast-evolving SNe (in gray).
Several NS-NS merger models, scaled to a distance of 40 Mpc, are shown as comparison from \citealt{1998ApJ...507L..59L} [LP98]; \citealt{2010MNRAS.406.2650M} [Met10];  \citealt{2013ApJ...775...18B} [B$\&$K] and  \citealt{2013MNRAS.430.2121P} [Piran et al]. 
{\bf Left panel:} We show the detection limits in the position of DLT17ck in the 6 months before  GW170817 and an inset with the detected light curve}
\label{fig:lightcurve}
\end{center}
\end{figure*}

Our discovery magnitude $r=17.46\pm0.03\,\rm{mag}$  at the distance of 39.5$\pm$2.6 Mpc  
\citep[distance modulus, $\mu$=32.98$\pm$0.15 mag using the Tully-Fisher relation][]{2001ApJ...553...47F} and Milky Way reddening $E(B-V)=0.109\,\rm{mag}$ \citep{2011ApJ...737..103S} brings DLT17ck to an absolute magnitude of $M_{r}=-15.8\pm0.1\,\rm{mag}$. This magnitude is consistent with that typically observed in faint CC SNe \citep{2014MNRAS.439.2873S} and brighter than that of some kilonova models proposed so far. However, in the hours after the discovery, it became clear that DLT17ck was a unique event. DLT17ck was indeed  cooling down and getting dimmer,  much faster than any other SN we ever observed. About 35 hours after GW170817, DLT17ck had dimmed by almost a magnitude \citep{2017GCN..21579....1A}.  By day 5, DLT17ck was already $\sim$ 4 magnitudes fainter than at discovery and disappeared below our magnitude limit the day after. At the same time, DLT17ck remained detectable in the near-infrared for a longer time. In Figure \ref{fig:lightcurve} (right panel), we compare the DLT40 light curve of DLT17ck with those of the most rapid transients available in the literature. DLT17ck evolves faster than any other known SN (gray points) and peaked probably between our discovery images and our third detection (respectively 11 and 35 hours after GW170817)\footnote{The possibility that DLT17ck is not related to GW170817, and exploded prior to the event, is discussed in Sec.\ref{sec:prediscovery}}.

Regardless of the energy source powering them, the light curves of astronomical transients like supernovae and kilonovae are regulated by the same physics. At early times, the photons released can not immediately escape due to the high optical depth. The photon \emph{diffusion time} depends on the ejecta mass, the opacity and the ejecta velocity \citep{1982ApJ...253..785A}. For kilonovae, the ejected mass has been predicted to be between 10$^{-4}$ and 10$^{-2}$ \Msun{} depending on the lifetime of the hypermassive neutron star that forms at the moment of coalescence.  A longer lifetime corresponds to a larger ejected mass and a brighter and longer-lasting optical electromagnetic counterpart \citep{2013ApJ...774...25K, Metzger17}.

Because of the high neutron fraction the nucleosynthesis in the ejected material is driven by the r-process, producing a significant fraction of lanthanide that dominates the opacity. Because of a large uncertainty in lanthanide opacity,  the ejecta opacity is not well constrained;  it should be between 1 and 100 $\rm cm^2$ $\rm g^{-1}$  \cite[closer to 1 for ejecta with a small amount of lanthanide elements; ][]{Metzger17}. Finally, velocities in the range 0.1-0.3 times the speed of light are also expected \citep[see ][ and reference therein]{Metzger17}. Using equation 5 from \cite{Metzger17}, 

\begin{eqnarray}
t_{peak}  \equiv (\frac{3Mk}{4\pi \beta v c})^{1/2} \approx 1.6 d (\frac{M}{10^{-2} M_{\odot}})^{1/2}  (\frac{v}{0.1c})^{-1/2} \nonumber \\
  (\frac{k}{1 cm^{2} g^{-1}})^{1/2} 
 \end{eqnarray}
 
\noindent where $\beta$~$\approx$ 3, $M$ is the ejected mass, $v$ is the expansion velocity, $k$ the opacity and $t_{peak}$ is the time of the peak,  we can give a a rough estimate of the ejected mass.  
Soon after our first detection (11 hours after explosion), a few groups reported a flattening or slightly increase of the luminosity \citep{2017GCN..21560....1A, 2017GCN..21565....1A}, but our second detection (35 hours after explosion) shows the object fading.  We then assume Aug. 18.528 UT  (24 hours after  GW170817) as the epoch of the peak. We use an opacity of 1-10 $\rm cm^{2}/g$ since the early blue peak should not contain large amounts of lanthanide \citep{Metzger17} and an expansion velocity of 0.2 $\times$ c. With these values, we obtain an ejected mass of $\approx$ 3 $\times$ 10$^{-3}$  - 10$^{-2}$   \Msun{}. However, the equation we used is an approximation and more careful models are needed.
Comparing the DLT40 light curve with several kilonova models (see Figure \ref{fig:lightcurve}), we found two models evolving as fast as DLT17ck which we describe below: The model by  \citealt{2010MNRAS.406.2650M} (Met10) assumes a radioactive powered emission and an ejected mass of $10^{-2}$ M$_{\odot}$, outflow speed of $v=0.1c$ and iron like opacity; the model by \citealt{2013ApJ...775...18B} (B$\&$K) assumes an ejected mass of $10^{-3}$ M$_{\odot}$, velocity of 0.1 c and a typical lanthanide opacity. Both models are consistent with the ejected mass we computed above, and support the kilonova interpretation.

\begin{figure*}
\begin{center}
\includegraphics[width=16cm]{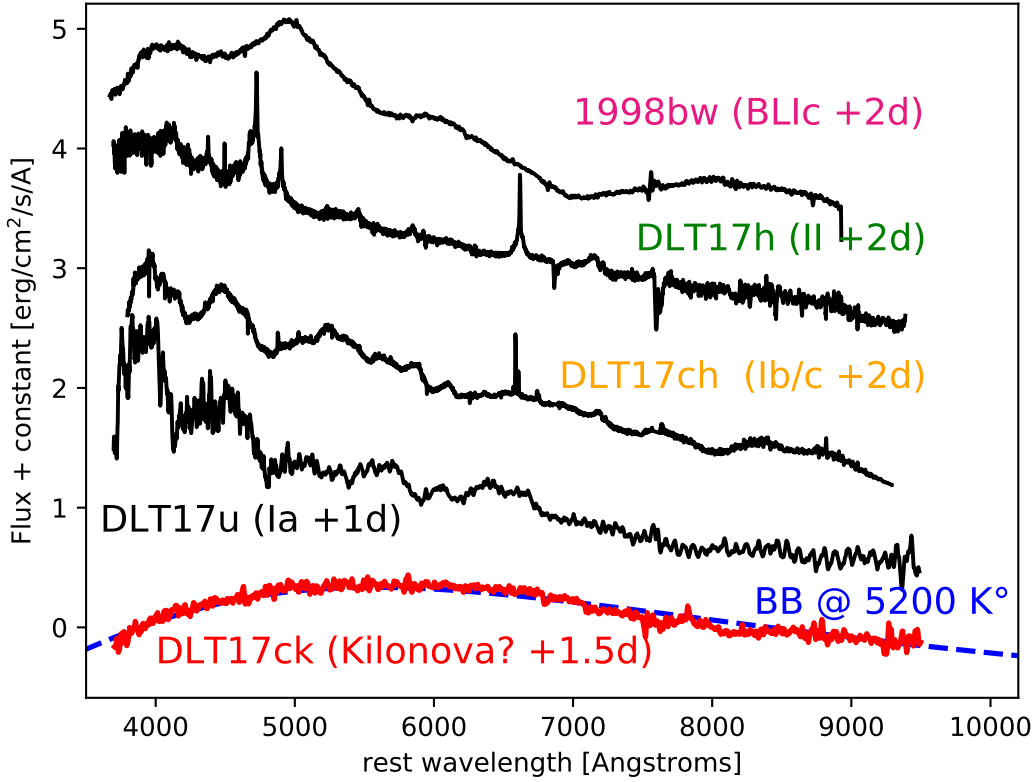}
\caption{DLT17ck spectrum at 35 hours after the GW170817 compared with spectra of young SNe at similar epochs. DLT17ck is cooling much 
faster than any previously observed explosive transient. A blackbody fit indicates a temperature of $\approx$5200 K. Data from: DLT17u (FLOYDS), 
DLT17ch (SALT), DLT17h (SALT), DLT17ck (NTT), SN1998bw (Danish 1.54 telescope + DFOSC). 
The presence of an emission feature at $\sim$7800 $\rm \AA$ is suspicious due to the presence of telluric lines close its position.}
 \label{fig:spectrum}
\end{center}
\end{figure*}

Further evidence for the kilonova hypothesis comes from the analysis of DLT17ck spectra. Spectroscopic observations were performed by \cite{2017GCN..21547....1A}  about 12 hrs after GW170817, showing a blue and featureless continuum. This supports the idea that DLT17ck was discovered young, although a blue and featureless continuum is also common for young SNe II and GRB afterglows. The fast cooling of DLT17ck (and hence the small ejected mass) became evident as more spectra were collected. The extended-Public ESO Spectroscopic Survey for Transient Objects \citep[ePESSTO ][]{2015A&A...579A..40S} observed DLT17ck $\sim$ 35 hours after GW170817, reporting a featureless spectrum, with a much redder continuum than that observed in SN spectra at similar phases \citep[][ see Figure \ref{fig:spectrum}]{2017GCN..21582....1A}. A black-body fit to the spectrum revealed  a temperature of $\approx 5200$\,K. Considering a spherically symmetric explosion and a black body emission, the radius of the kilonova should have expanded from the radius of a neutron star (few tenth 10$^{5}$ cm ) to $\sim$ 7.3 $\times$ 10$^{14}$ cm. Under homologous expansion this requires a velocity expansion of 0.2\,$c$.

\section{Search for pre-discovery outbursts in historical data.}
\label{sec:prediscovery}
In the standard kilonova model, we only expect a bright electromagnetic signature after coalescence. We can test this by looking at DLT40 observations taken before 2017 August 17. NGC~4993 is one of the galaxies monitored by the DLT40 supernova search, observed on average every 3 days from February 2017 to July 2017 (see Table \ref{tab1}).  Our images show no sign of an optical transient down to a limit of $m_{r} \sim 19$ mag (see Figure 3), corresponding to $M_{r} \sim -14$ mag at the adopted distance of NGC~4993. Similarly, the field was also observed from 2013 to 2016 from La Silla QUEST on the ESO 1.0 meter telescope with no detection to a limit of $R \sim$ 18 mag \citep{2017GCN..21599....1A}.

The last DLT40 non-detection at the position of DLT17ck is on 2017 July 27th (21 days before the LVC event) down to $m_{r} = 19.1$ mag. Combining this limit with the extremely fast timescale of the transient, its blue continuum in the early spectra, its rapid cooling, and its photometric consistency with some kilonova models makes it extremely unlikely DLT17ck can be explained by any kind of supernova unrelated to the GW/GRB event. Rather, all the evidence favors that DLT17ck was discovered young, and is the optical counterpart of GW170817 and GRB 524666471/170817529. 

\section{Summary and Future Prospects}\label{sec:summary}
In this paper, we presented the discovery of DLT17ck in the error region of the LVC event GW170817 and the \textit{Fermi} short GRB 524666471/170817529. DLT17ck is  characterized by a very fast optical evolution, consistent with some kilonova models and with a small ejecta mass (10$^{2}$ - 10$^{-3}$ \Msun{}). Spectroscopic observations conducted about 35 hrs after the explosion show a featureless continuum with blackbody temperature of 5200 K, confirming the fast evolution of DLT17ck compared to the evolution of other transients like classical supernovae.
In addition, it is also surprising that at such a low temperature, no features are visible. We may speculate that this is the result of blending due to the high velocity of the expanding ejecta. Given the coincidence with the LVC event and the short \textit{Fermi} GRB, it is likely the optical counterpart of the merging of two neutron stars in a binary system. This event represents a milestone for astronomy, being the first multimessenger event from which both photons and gravitational waves have been detected.

The unprecedented characteristics of DLT17ck raise a question as to the rates of such objects. The daily cadence of the DLT40 search can help constrain the rates of kilonovae and other rapidly-evolving transients. Details of rate measurements will be presented in a dedicated paper (Yang et al., in preparation), while here we report some of the results related to kilonovae. Using the galaxies within 40 Mpc that we have observed in the last two years, and under the simplifying assumption that all kilonovae have a light curve similar to DLT17ck, we find an upper limit (at 95$\%$ confident level)  to the rate of kilonovae of 
$\rm 0.48_{-0.15}^{+0.9}$ binary neutron stars (BNS) SNu\footnote{SNu = $(100~yr)^{-1}$ $ (10^{10}  L^{B}_{\odot})^{-1} $ }. For a Milky Way luminosity $\sim 2 \times 10^{10} \; L_\odot$, this translates to an upper limit of 9 Galactic kilonovae per millennium. This limit is not too stringent since it is two orders of magnitude larger than the Galactic rate of binary neutron star coalescence of 24 Myr$^{-1}$ estimated by \cite{2015MNRAS.448..928K} from known neutron-star binaries. 

We can convert our luminosity-based kilonova rate to a volumetric rate using the local luminosity density from \cite{2003ApJ...592..819B}. This gives a limit of $\rm 9.4 \pm 0.8 \times 10^{-5}$ kilonovae $\rm Mpc^{-3}$ $\rm yr^{-1}$. This is consistent with previous limits \citep[$<$0.05 $\rm Mpc^{-3}$ $\rm yr^{-1}$; ][]{2013ApJ...779...18B}, that however were based on hypothetical parameters for the BNS optical light curve, and is comparable to the volumetric rate of fast optical transients, 4.8 - 8.0 $\rm  \times 10^{-6}$  $\rm Mpc^{-3}$ $\rm yr^{-1}$ found by \cite{2014ApJ...794...23D}. 

Looking forward to the O3 LVC run, in 2018, it is useful to explore strategies to detect EM counterparts of NS-NS mergers. DLT17ck was discovered independently by several groups (eg. SWOPE and DLT40; \citealt{2017GCN..21529....1A} and  \citealt{2017GCN..21531....1A}),  using the approach of targeting nearby galaxies within the LVC region with small field-of-view instruments \citep{2016ApJ...820..136G}. Several wide-field searches were also able to identify the transient \citep{2017GCN..21530....1A, 2017GCN..21542....1A, 2017GCN..21546....1A, 2017GCN..21553....1A}, but only after reports from the targeted searches. This was likely due to the challenge of analyzing a large amount of data in a short period of time. 

The small field-of-view strategy, and certainly our discovery, was successful because GW170817/DLT17ck was extremely nearby. The short \textit{Fermi} GRB associated with DLT17ck is the closest ever discovered \citep[see ] [ for a review of short GRBs]{2014ARA&A..52...43B}. However, with the expected increase in sensitivity of the LVC detectors, in O3 the volume where NS-NS mergers can be detected will reach 150 Mpc, increasing further to 200 Mpc at full sensitivity \citep[2019+; ][]{2016LRR....19....1A}. At these distances galaxy catalogs are incomplete  \citep{2016ApJ...827L..40S} and the sheer number of galaxies will likely favor wide-field strategies. Nonetheless, because the Virgo horizon distance during O3 is predicted to be $65-115$\,Mpc \citep{2016LRR....19....1A}, the small field-of-view strategy may still be important for the best-localized sources. DLT40 reaches a limiting magnitude of $r \sim 19$ mag in 45 to 60 second exposures. Taking a more conservative limit of 18.5 mag, we would expect to be able to see sources like DLT17ck out to 70 Mpc. Increasing the exposure time to reach a depth of $\sim$ 20 mag would allow us to observe binary neutron star mergers in the full range of the Virgo interferometer.

\begin{table*}
\centering
 \begin{minipage}{140mm}
\caption{Photometric Data for DLT17ck}
\scriptsize
\begin{tabular}{ccccc|ccccc}
\hline
Date & JD & mag$^{a}$ $^{b}$ &  Filter$^{c}$ & telescope &Date & JD & mag$^{a}$ $^{b}$&  Filter$^{c}$ & telescope\\
\hline
 2017-08-17   &  2457983.493  &   17.46   0.03  &  $r$     &  Prompt 5  &  2017-03-07   &  2457819.772  &  $>$ 20.90    &  $r$   &    Prompt 5 \\
 2017-08-18   &  2457983.528  &   17.56   0.04  &  $r$     &  Prompt 5  &  2017-03-10   &  2457822.595  &  $>$ 19.97    &  $r$   &    Prompt 5 \\
 2017-08-18   &  2457984.491  &	  18.00   0.06  &  $r$     &  Prompt 5  &  2017-03-11   &  2457823.592  &  $>$ 19.37    &  $r$   &    Prompt 5 \\
 2017-08-19   &  2457984.510  &   18.29   0.06  &  $r$     &  Prompt 5  &  2017-03-12   &  2457824.594  &  $>$ 19.39    &  $r$   &    Prompt 5 \\
 2017-08-19   &  2457985.476  &   19.34   0.08  &  $r$     &  Prompt 5  &  2017-03-13   &  2457825.586  &  $>$ 19.20    &  $r$   &    Prompt 5 \\
 2017-08-19   &  2457985.478  &   19.29   0.12  &  $r$     &  Prompt 5  &  2017-03-26   &  2457838.881  &  $>$ 20.37    &  $r$   &    Prompt 5 \\
 2017-08-21   &  2457986.503  &   20.18   0.10  &  $r$     &  Prompt 5  &  2017-03-27   &  2457839.714  &  $>$ 21.14    &  $r$   &    Prompt 5 \\
 2017-08-22   &  2457987.504  &	  20.92   0.12  &  $r$     &  Prompt 5  &  2017-03-28   &  2457840.717  &  $>$ 20.86    &  $r$   &    Prompt 5 \\
 2017-07-27   &  2457961.599  &	 $>$ 19.84      &  $r$     &  Prompt 5  &  2017-03-29   &  2457841.720  &  $>$ 21.03    &  $r$   &    Prompt 5 \\
 2017-07-27   &  2457962.495  &	 $>$ 19.36      &  $r$     &  Prompt 5  &  2017-03-30   &  2457842.666  &  $>$ 20.74    &  $r$   &    Prompt 5 \\
 2017-05-15   &  2457888.762  &  $>$ 19.79      &  $r$     &  Prompt 5  &  2017-03-31   &  2457843.713  &  $>$ 20.83    &  $r$   &    Prompt 5 \\
 2017-05-16   &  2457889.751  &	 $>$ 19.88      &  $r$     &  Prompt 5  &  2017-04-02   &  2457845.704  &  $>$ 20.90    &  $r$   &    Prompt 5 \\
 2017-05-17   &  2457890.796  &  $>$ 19.61      &  $r$     &  Prompt 5  &  2017-04-04   &  2457847.695  &  $>$ 20.73    &  $r$   &    Prompt 5 \\
 2017-05-20   &  2457893.500  &  $>$ 19.88      &  $r$     &  Prompt 5  &  2017-04-05   &  2457848.700  &  $>$ 20.87    &  $r$   &    Prompt 5 \\
 2017-05-21   &  2457894.562  &  $>$ 20.27      &  $r$     &  Prompt 5  &  2017-04-06   &  2457849.857  &  $>$ 20.63    &  $r$   &    Prompt 5 \\
 2017-05-22   &  2457895.545  &  $>$ 20.65      &  $r$     &  Prompt 5  &  2017-04-07   &  2457850.699  &  $>$ 20.24    &  $r$   &    Prompt 5 \\
 2017-05-28   &  2457901.715  &  $>$ 19.69      &  $r$     &  Prompt 5  &  2017-04-08   &  2457851.695  &  $>$ 19.74    &  $r$   &    Prompt 5 \\
 2017-05-29   &  2457902.548  &  $>$ 20.16      &  $r$     &  Prompt 5  &  2017-04-09   &  2457852.679  &  $>$ 19.56    &  $r$   &    Prompt 5 \\
 2017-05-30   &  2457903.547  &  $>$ 20.12      &  $r$     &  Prompt 5  &  2017-04-13   &  2457856.861  &  $>$ 19.45    &  $r$   &    Prompt 5 \\
 2017-05-31   &  2457904.544  &  $>$ 20.70      &  $r$     &  Prompt 5  &  2017-04-14   &  2457857.678  &  $>$ 18.41    &  $r$   &    Prompt 5 \\
 2017-06-01   &  2457905.542  &  $>$ 20.45      &  $r$     &  Prompt 5  &  2017-04-15   &  2457858.666  &  $>$ 20.07    &  $r$   &    Prompt 5 \\
 2017-06-02   &  2457906.511  &  $>$ 20.24      &  $r$     &  Prompt 5  &  2017-04-21   &  2457864.654  &  $>$ 21.01    &  $r$   &    Prompt 5 \\
 2017-06-02   &  2457907.498  &  $>$ 20.06      &  $r$     &  Prompt 5  &  2017-04-22   &  2457865.642  &  $>$ 20.93    &  $r$   &    Prompt 5 \\
 2017-06-19   &  2457923.645  &  $>$ 20.61      &  $r$     &  Prompt 5  &  2017-04-23   &  2457866.760  &  $>$ 21.02    &  $r$   &    Prompt 5 \\
 2017-06-19   &  2457924.481  &  $>$ 19.65      &  $r$     &  Prompt 5  &  2017-04-24   &  2457867.652  &  $>$ 21.04    &  $r$   &    Prompt 5 \\
 2016-12-21   &  2457743.834  &  $>$ 20.44      &  $r$     &  Prompt 5  &  2017-04-26   &  2457869.631  &  $>$ 20.67    &  $r$   &    Prompt 5 \\
 2017-02-06   &  2457790.858  &  $>$ 21.39      &  $r$     &  Prompt 5  &  2017-04-27   &  2457870.626  &  $>$ 20.92    &  $r$   &    Prompt 5 \\
 2017-02-07   &  2457791.823  &  $>$ 21.34      &  $r$     &  Prompt 5  &  2017-04-28   &  2457871.622  &  $>$ 20.65    &  $r$   &    Prompt 5 \\
 2017-02-08   &  2457792.826  &  $>$ 21.26      &  $r$     &  Prompt 5  &  2017-04-29   &  2457872.694  &  $>$ 20.66    &  $r$   &    Prompt 5 \\
 2017-02-09   &  2457793.835  &  $>$ 21.10      &  $r$     &  Prompt 5  &  2017-04-30   &  2457873.618  &  $>$ 20.71    &  $r$   &    Prompt 5 \\
 2017-02-10   &  2457794.824  &  $>$ 20.58      &  $r$     &  Prompt 5  &  2017-05-01   &  2457874.615  &  $>$ 20.60    &  $r$   &    Prompt 5 \\
 2017-02-11   &  2457795.825  &  $>$ 20.33      &  $r$     &  Prompt 5  &  2017-05-03   &  2457876.665  &  $>$ 20.76    &  $r$   &    Prompt 5 \\
 2017-02-12   &  2457796.756  &  $>$ 19.90      &  $r$     &  Prompt 5  &  2017-05-04   &  2457877.594  &  $>$ 20.55    &  $r$   &    Prompt 5 \\
 2017-02-13   &  2457797.747  &  $>$ 20.16      &  $r$     &  Prompt 5  &  2017-05-05   &  2457878.606  &  $>$ 20.30    &  $r$   &    Prompt 5 \\
 2017-02-14   &  2457798.692  &  $>$ 19.85      &  $r$     &  Prompt 5  &  2017-05-06   &  2457879.577  &  $>$ 20.25    &  $r$   &    Prompt 5 \\
 2017-02-17   &  2457801.725  &  $>$ 20.37      &  $r$     &  Prompt 5  &  2017-08-25   &  2457990.504  &  $>$ 20.89    &  $r$   &    Prompt 5 \\
 2017-02-19   &  2457803.828  &  $>$ 20.83      &  $r$     &  Prompt 5  &  2017-08-26   &  2457991.504  &  $>$ 20.37    &  $r$   &    Prompt 5 \\
 2017-03-05   &  2457817.886  &  $>$ 20.78      &  $r$     &  Prompt 5  &  2017-08-26   &  2457992.489  &  $>$ 19.90    &  $r$   &    Prompt 5 \\
 2017-03-06   &  2457818.784  &  $>$ 20.91      &  $r$     &  Prompt 5  &     --        &    --         &   --          &  --    &    --       \\
 \hline
\end{tabular}
\label{tab1}
(a) : Data has not been corrected for extinction. \\
(b) : Limit magnitude are 5 $\sigma$ detection limit.\\\
(c) : $Open$ filter calibrated to $r$. \\
\end{minipage}
\end{table*}

\acknowledgments
Research by DJS and L.T. l is supported by NSF grant AST-1412504 and AST-1517649.
AC acknowledges support from the NSF award $\#$1455090.
The work of SY was supported by the China Scholarship Council(NO. 201506040044).

\end{document}